\documentclass[twocolumn,showpacs,preprintnumbers,amsmath,amssymb]{revtex4}

\usepackage{graphicx}
\usepackage{dcolumn}
\usepackage{bm}
\usepackage{amsthm}
\theoremstyle{definition}\newtheorem{Th}{Theorem}
\theoremstyle{definition}\newtheorem{Lm}{Lemma}

\begin{document}
\title{\bf{Sequential scheme for locally discriminating bipartite unitary operations without inverses}}
\author{Lvzhou Li}\email{lilvzh@mail.sysu.edu.cn
(L. Li).}%

 \affiliation{%
 Institute of Computer Science Theory, School of Data and Computer Science, Sun Yat-sen University, Guangzhou 510006,
 China
}%

\affiliation{%
 The Key Laboratory of Machine Intelligence and Advanced Computing (Sun Yat-sen University)
Ministry of Education, China
}%
\date{\today}

 \begin{abstract}
Local distinguishability of  bipartite unitary operations has recently received much attention.  A nontrivial and interesting question concerning this subject is whether there is a sequential scheme for locally discriminating between two  bipartite unitary operations, because a sequential scheme usually represents the most economic strategy for discrimination. An affirmative answer to this question was given  in the literature, however with two limitations: (i) the  unitary operations  to be discriminated were limited to act on $d\otimes d$, i.e., a two-qudit system, and (ii) the inverses of the unitary operations  were assumed to be  accessible, although this assumption may be unrealizable in experiment. In this paper, we improve the result by removing the two limitations. Specifically, we show that any two  bipartite unitary operations acting on $d_A\otimes d_B$ can be locally discriminated by a sequential scheme, without using the inverses of the unitary operations. Therefore, this paper enhances  the applicability and feasibility of the sequential scheme for locally discriminating unitary operations.

 \end{abstract}
  \maketitle

\section{ Introduction}

Distinguishability of unitary operations is a fundamental problem in
quantum information and has received extensive
attention. Discrimination of  unitary operations is generally transformed to discrimination of quantum states by preparing an input state and then discriminating the output states generated by different unitary operations. However, distinguishability of unitary operations shows some interesting properties essentially different from  that  of quantum states, especially in the case of multiple queries.

Two unitary operations $U$ and $V$ are said to be
perfectly distinguishable (with  a single query), if there
exists an input state $|\psi\rangle$ such that $U|\psi\rangle\perp
V|\psi\rangle$.   It has been  shown that  $U$ and
$V$ are perfectly distinguishable
if, and only if $\Theta(U^\dagger V)\geq\pi$, where $\Theta(W)$
denotes the length of the smallest arc containing all the
eigenvalues of $W$ on the unit circle \cite{Acin,Paris}. The situation changes dramatically when  multiple queries  are allowed, since any two different unitary operations are perfectly distinguishable in this case. Specifically,
it was shown that for any two different unitary
operations $U$ and $V$, there exist a finite number $N$ and a
suitable state $|\varphi\rangle$ such that $U^{\otimes
N}|\varphi\rangle\perp V^{\otimes N}|\varphi\rangle$  \cite{Acin,Paris}.
Such a discriminating scheme is intuitively called a {\it parallel scheme}. Note that in the parallel scheme, an $N$-partite
entangled state as an input is required  and plays a crucial role. Then, the
result was further refined in \cite{Duan1} by showing that the
entangled input state is not necessary for perfect discrimination of unitary operations. Specially, \cite{Duan1} showed
that for any two different unitary operations $U$ and $V$, there
exist an input state $|\varphi\rangle$ and auxiliary unitary operations
$w_1,\dots,w_N$ such that $Uw_NU\dots w_1U|\varphi\rangle\perp
Vw_NV\dots w_1V|\varphi\rangle$. Generally, such
a discriminating scheme is called a {\it sequential scheme}.

Note that in these researches mentioned above, it was assumed by default that the  unitary operations to be discriminated are under the
complete control of a single party who can  perform any physically
allowed operations to achieve an optimal discrimination. Actually, a more complicated case is that the
 unitary operations to be discriminated   are shared by
several spatially separated parties. Then, in this case a reasonable constraint
on the discrimination is that each party can only make local
operations and classical communication (LOCC). Despite this constraint,
Refs. \cite{Duan2, Zhou} independently showed that any two bipartite unitary
operations can be perfectly discriminated by LOCC when   a finite
number  of queries are allowed. This implies that LOCC reaches distinguishability that global operations would have,  probably with more queries to the unitary operations.

\begin{figure}
 \includegraphics{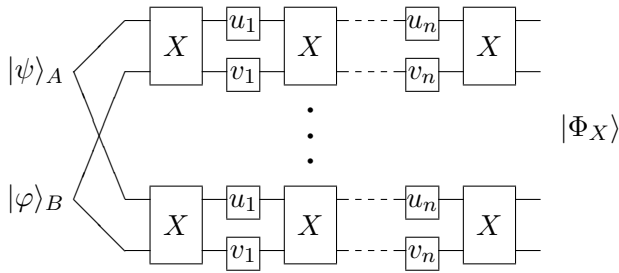}
\caption{ A mixed scheme for perfectly discriminating bipartite unitary
operations $X\in\{U,V\}$ by LOCC. $u_i$ and $v_i$ are single-particle unitary operations. $|\Phi_X\rangle$ denotes the output state of the circuit. A perfect discrimination between $U$ and $V$ means that there exists an input state $|\psi\rangle_A|\varphi\rangle_B$ such that the output states $|\Phi_X\rangle$ corresponding to different $X$ are orthogonal (that is, $|\Phi_U\rangle\perp |\Phi_V\rangle $),   and then  $|\Phi_U\rangle$ and $|\Phi_V\rangle$ can be perfectly discriminated by LOCC \cite{Walgate}.  Note that one of  $|\psi\rangle_A$ and $|\varphi\rangle_B$ held by Alice and Bob, respectively, must be a multipartite entangled state. } \label{Fig 1}
\end{figure}

It is worth mentioning that  distinguishability of unitary operations has interesting relations with other issues. For instance,  it has a closed relation with  universality of quantum gates \cite{Bry} as shown in \cite{Zhou}, and it is also related to the analysis of
numerical range \cite{Horn} as presented in \cite{Duan2}. Despite different methods used in \cite{Duan2, Zhou}, the main idea of them, which is depicted in Fig. \ref{Fig 1}, can be roughly described as follows.

(i) For two bipartite unitary operations $U$ and $V$ shared
by Alice and Bob that satisfy certain conditions,  there exist a finite number $N$ and a product
state $|\psi\rangle_A|\varphi\rangle_B$, such that $U^{\otimes
N}|\psi\rangle_A|\varphi\rangle_B\perp V^{\otimes N}|\psi\rangle_A|\varphi\rangle_B$, where $|\psi\rangle_A$
and $|\varphi\rangle_B$ are two N-partite states prepared by Alice and
Bob, respectively, and one of which must be an N-partite entangled
state. Such an entangled state held by
one party is called {\it local entanglement}.

(ii) For any two general bipartite unitary
operations $U$ and $V$, one can construct a
quantum circuit  $f(X)=Xw_1X\dots w_nX$ with $X\in\{U,V\}$  and
 a  sequence of local unitary operations
$w_1,\dots,w_n$ (each $w_i$ has the form $w_i=u_i\otimes
v_i$), such that $f(U)$ and $f(V)$ satisfy the desired condition
stated in item (i). Thus $f(U)$ and $f(V)$ can be discriminated as
in item (i), which means that $U$ and $V$ can be perfectly
discriminated by LOCC.

In the above procedure, there generally needs to be a {\it mixed scheme} which combines the sequential and the parallel schemes to
achieve a perfect discrimination. At the same time, one of the two parties who share the bipartite unitary operations
 must prepare  a multipartite entangled state.
But, note that a sequential scheme usually represents the most economic strategy for discrimination, since it does not require entanglement as indicated by the sequential scheme \cite{Duan1} compared with the parallel scheme \cite{Acin,Paris}.
Then a natural question, as proposed in \cite{Duan2},  is whether there is a sequential scheme  for perfectly discriminating   bipartite unitary operations by LOCC.

In Ref. \cite{Li08} we answered the above question affirmatively by proving that
any two bipartite unitary operations acting on  $d\otimes d$
(i.e., a two-qudit system), in
principle, can be perfectly discriminated  by LOCC  with a
sequential scheme, when a finite number of queries are allowed.
However, there is still room for improvement  at least from the following two aspects.

First, the result only applies to the unitary operations acting on $d\otimes d$, where the two subsystems have the same dimension.  Then, how about the general unitary operations acting on $d_A\otimes d_B$ with $d_A\neq d_B$?

Second, in the proof of the result,   in order to discriminate $U$ and $V$, their inverses $U^\dagger$ and $V^\dagger$ were assumed to be accessible as long as $U$ and $V$ are accessible. This assumption is also fundamental in \cite{Zhou}. One may  think that  a unitary operation $U$ can   be regarded as a black box with
input and output ports, and
then the inverse  $U^\dagger$ can be obtained   by simply reversing the whole setup. However, by the current experiment technology, such an operation may not be easily realized
or even cannot be realized. Then, a natural question is, can we avoid using the inverse $U^\dagger$?  The answer was shown to be ``yes'' for the case of $d=2$ in \cite{Zhou, Li08}, but it was not clear for the case of higher dimensions.

Therefore, in this paper we improve the result of \cite{Li08} by considering the above two points. Specifically, we show that any two  different bipartite unitary operations acting on $d_A\otimes d_B$, allowed to be queried a finite number of times,  can be locally discriminated by a sequential scheme, without using the inverses of the unitary operations. This result rests on universality of quantum gates \cite{Har09}.

The rest of this paper is organized as follows. Section \ref{Pre} presents some preliminaries of this paper. The main result   is presented in Section \ref{Main}. A conclusion is made in Section \ref{Con}.

\section {preliminaries}\label{Pre}

We first recall a  result regarding
the distinguishability of unitary operations in \cite{Duan1}.

\begin{Lm}\label{Lm1}Let $U$ and $V$ be two different unitary
operations. Then there exist a finite number $N$, auxiliary unitary operations
$w_1,\dots,w_{N}$, and an input state $|\psi\rangle$ such that
\begin{align*}
Uw_{N}U\dots w_1U|\psi\rangle\perp Vw_{N}V\dots
w_1V|\psi\rangle.
\end{align*}
\end{Lm}
The above scheme is the so-called sequential scheme for
discriminating two unitary operations. Also, there was a parallel scheme \cite{Acin,Paris} which claims that for any two different unitary
operations $U$ and $V$, there exist a finite number $N$ and a
 state $|\varphi\rangle$ such that $U^{\otimes
N}|\varphi\rangle\perp V^{\otimes N}|\varphi\rangle$.

In this paper, we  focus on   unitary operations acting on a bipartite system $AB$.  Assume that each subsystem $X$ ($X\in\{A, B\}$) has a $d_X$-dimensional Hilbert space ${\cal H}_X$. Then the whole state space of  $AB$ is  ${\cal H}_A\otimes{\cal H}_B$, and we will use $d_A\otimes d_B$ as an abbreviation  for it.
Let ${\cal U}(d_A\otimes d_B)$ denote the set of all  unitary operations acting on $d_A\otimes d_B$, and let ${\cal U}_d$ denote the set of all unitary operations acting on a $d$-dimensional Hilbert space.
$U\in {\cal U}(d_A\otimes d_B)$ is said to be {\it imprimitive } if there exist $|\varphi\rangle\in {\cal H}_{A}$ and $|\phi\rangle\in {\cal H}_{B}$ such that the state $U|\varphi\rangle|\phi\rangle$ is entangled. Otherwise, it is primitive. Equivalently, as mentioned in \cite{Bry,Har09}, $U$ is primitive if it can be written as $U_A\otimes U_B$ when  $d_A\neq d_B$,  or can be written as $U_A\otimes U_B$ or $(U_A\otimes U _B)P$ when $d_A=d_B$,  where $U_A\in {\cal U}_{d_A}$, $U_B\in {\cal U}_{d_B}$, and
$P$ is a swapping operation, i.e.,
$P|x\rangle|y\rangle=|y\rangle|x\rangle$.

 Let ${\cal P}$ denote a subset of ${\cal U}(d_A\otimes d_B)$ as
 $${\cal P}\equiv\{U: U=U_A\otimes U_B, U_A\in {\cal U}_{d_A}, U_B\in {\cal U}_{d_B}\}.$$

Harrow \cite{Har09} obtained a result concerning  universality of quantum gates as follows.

\begin{Lm} ${\cal P}$ together with an imprimitive $V$ can generate
any  unitary operation acting on $d_A\otimes d_B$. More specifically, there exists an integer $N$ such that for any $U\in {\cal U}(d_A\otimes d_B)$ there is $U=X_N\cdots X_2 X_1$ for  $X_i\in {\cal P}\cup\{V\}$ with $i=1,\cdots, N$.\label{Lm2}
\end{Lm}

The above result improves the one in \cite{Bry}, since  the inverse of $V$ is not used in the above, whereas it was required in \cite{Bry}. The result will be a base of  this paper. The following technical lemma is also required in order to proving the main result of this paper. A detailed proof of the lemma for the case of $d_A=d_B$ was given in \cite{Li08}, and one can easily extend the proof to the general case of $d_A\neq d_B$.

\begin{Lm} For  unitary operation $U\in {\cal U}(d_A\otimes d_B)$,
$U^\dagger=WUW^\dagger$ holds for all $W\in {\cal S}\equiv\{(\sigma_z\oplus I)\otimes I, (\sigma_y\oplus I)\otimes
I,I\otimes(\sigma_z\oplus I),I\otimes(\sigma_y\oplus I)\}$ if, and
only if $U$ has the form $U=e^{ixu_1\otimes u_2}$ for some real
number $x$, where $u_1=\sigma_x\oplus 0_{(d_A-2)}$, $u_2=\sigma_x\oplus 0_{(d_B-2)}$,  with
$\sigma_x$, $\sigma_y$ and $\sigma_z$ being Pauli operators. \label{Lm3}
\end{Lm}

\section {Sequential scheme for local discrimination without inverses}\label{Main}
Now, we are  in a position to  give our main result.  We show that any two different unitary  operations  acting on $d_A\otimes d_B$, allowed with a finite number of queries, can be locally discriminated by a sequential scheme without using the inverses. The result is  depicted in  Fig \ref{Fig 2}, and formally presented in Theorem \ref{th1} below. In the rest of this paper, we will use the  notation $f(X)$ to denote a sequential circuit of the form depicted in Fig \ref{Fig 2}.

\begin{figure}
 \includegraphics[width=9cm]{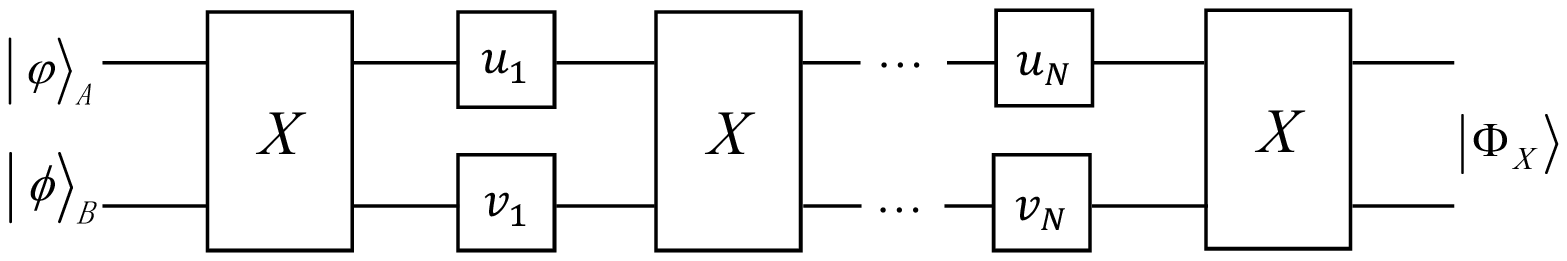}
\caption{ A sequential scheme for locally discriminating unitary operations $U$ and $V$ acting on $d_A\otimes d_B$. Here $X$ represents the unknown unitary operation $U$ or  $V$, $N+1$ is the finite number of applying $X$, $\{u_i\}_{i=1}^{N}$ and $\{v_i\}_{i=1}^{N}$ are unitary operations acting on ${\cal H}_{A}$ and ${\cal H}_{B}$, respectively, and $|\varphi\rangle_A|\phi\rangle_B$ is the input state. The output state $|\Phi_U\rangle$  and $|\Phi_V\rangle$ are orthogonal, and thus can be perfectly discriminated by LOCC \cite{Walgate}.} \label{Fig 2}
\end{figure}

\begin{Th}\label{th1} For any two different  operations $U,
V$  acting on $d_A\otimes d_B$, there exist two finite sequences of unitary operations $\{u_i\}_{i=1}^{N}\subseteq {\cal U}_{d_A}$ and  $\{v_i\}_{i=1}^{N}\subseteq {\cal U}_{d_B}$ , and a product state $|\varphi\rangle|\phi\rangle\in d_A\otimes d_B$
such that
\begin{align*}
U(u_{N}\otimes v_{N})\cdots U(u_{1}\otimes v_{1})U|\varphi\rangle|\phi\rangle\\
\perp V(u_{N}\otimes v_{N})\cdots V(u_{1}\otimes v_{1})V|\varphi\rangle|\phi\rangle.
\end{align*}
\end{Th}

{\noindent\bf Remark.} The above result improves the one in \cite{Li08} in two aspects. First,  the inverses $U^\dagger$ and $V^\dagger$ are not used here,  whereas they were  required  in \cite{Li08} as well as in \cite{Zhou}. Actually,  it is  not easy to obtain $U^\dagger$  from $U$ in experiment.  Second, the result here holds for the general case of  $d_A\neq d_B$, but it was required that $d_A=d_B$ in \cite{Li08}.

 {\noindent\bf Proof of Theorem 1.} We prove the result by considering three cases: (i) both $U$ and $V$ are  primitive, (ii) one of them is primitive, and (iii) neither of them is primitive.

{\noindent\bf Case (i):  Both $U$ and $V$ are primitive.} Then it  suffices to
consider the  following three subcases.

Case (i-a): $U=U_A\otimes U_B$ and $V=V_A\otimes V_B$. Without
loss of generality, assume that $U_A\neq  V_A$. Then by Lemma \ref{Lm1},
$U_A$ and $V_A$ can be discriminated sequentially, and thus $U$ and $V$ can be  locally discriminated by a sequential scheme as in Fig \ref{Fig 2}.

Case (i-b): $U=U_A\otimes U_B$ and $V=(V_A\otimes V_B)P$. Note that this case occurs only  if $d_A=d_B$.  In this case,  it is easy to locally discriminate $U, V$ by applying them once, since by letting $|\Phi_X\rangle= X |\varphi\rangle|\phi\rangle$ with $X\in \{U,V\}$, we have
\begin{align*}
\langle\Phi_U|\Phi_V\rangle=\langle\varphi|U_A^\dagger
V_A|\phi\rangle\langle\phi|U_B^\dagger U_B|\varphi\rangle,
\end{align*}
which can be zero by setting $|\phi\rangle=V_A^\dagger
U_A|\varphi^\perp\rangle$.

Case (i-c): $U=(U_A\otimes U_B)P$ and $V=(V_A\otimes V_B)P$.  This case also occurs  only if $d_A=d_B$.
Without loss of generality, assume that $U_A\neq V_A$. Let
$f(X)=X(u\otimes v)X$,
 where $X\in \{U, V\}$ and $u, v$ are two given single-particle unitary operations. Then it is straightforward to  get that
\begin{align*}
f(U)=U_AvU_B\otimes U_BuU_A, \nonumber\\
f(V)=V_AvV_B\otimes V_BuV_A.
\end{align*}
It can be found that there always exists  $v$ such that $U_AvU_B\neq V_AvV_B$. By contradiction, suppose  $U_AvU_B=V_AvV_B$ holds for all $v$. Then we have $vU_BV_B^\dagger =U_A^\dagger V_Av$ for all $v$, which holds only if  $U_A=V_A$ and $U_B=V_B$.  This contradicts the premise that $U,V$ are different.

Therefore,  $f(U)$ and
$f(V)$ can be locally discriminated by a sequential scheme as in subcase (i-a), and so for $U$ and $V$.

{\noindent \bf Case (ii): One of $U$ and $V$ is primitive.} Without loss of
generality, assume that $V$ is primitive. We have the following discussion.

Case (ii-a):  $U$ is imprimitive and $V=V_A\otimes V_B$. By Lemma \ref{Lm2}, we can construct a sequential circuit $f(X)$ consisting of  local unitary  operations  and $X\in\{U,V\}$, such that \begin{align}f(U)=P_A\otimes
U_B+P'_A\otimes U'_B, \label{cq}\end{align}  where $U_B\neq U'_B$,
and $P_A$ and ${P}'_A$ are two projectors satisfying $P_A+{P}'_A=I_A$.
In other words, $f(U)$ is a controlled unitary transformation. At
the same time, it is clear that $f(V)\in {\cal P}$. Thus we let
$f(V)=V_A^{'}\otimes V_B^{'}$. Note that it holds that either $V_B^{'}\neq U_B$ or  $V_B^{'}\neq U'_B$. Without loss of generality, assume that $V_B^{'}\neq U'_B$.
Then as shown below, local discrimination between
$f(U)$ and $f(V)$  can be reduced to discrimination between $V_B^{'}$ and $ U'_B$.

Let $|\alpha\rangle_A\in{\cal H}_{A}$
satisfy $P_A^{'}|\alpha\rangle_A=|\alpha\rangle_A$. Then for any
$|\phi\rangle_B\in{\cal H}_{B}$, and $w_1,\dots, w_N$ acting on ${\cal H}_{B}$, we have
\begin{align*}
&f(U)(I\otimes w_N)f(U)\dots(I\otimes
w_1)f(U)|\alpha\rangle_A|\varphi\rangle_B\\=&|\alpha\rangle_A\otimes(U_B^{'}w_NU_B^{'}\dots
w_1U_B^{'})|\phi\rangle_B
\end{align*}
and
\begin{align*}
&f(V)(I\otimes w_N)f(V)\dots(I\otimes
w_1)f(V)|\alpha\rangle_A|\phi\rangle_B\\=&{V'_A}^{ N}|\alpha\rangle_A\otimes(V_B^{'}w_NV_B^{'}\dots
w_1V_B^{'})|\phi\rangle_B
\end{align*}
According to Lemma \ref{Lm1},  there exist a state
$|\phi\rangle_B$ and unitary operations $w_1,\dots, w_N$ such that the two output states  of  $B$ in the above are orthogonal.
Therefore, $f(U)$ and $f(V)$ can be locally discriminated by a sequential scheme, and so for $U$ and $V$.

Case (ii-b): $U$ is imprimitive and $V=(V_A\otimes V_B)P$.  As we did in subcase (ii-a), construct a sequential circuit $f(X)$ such that $f(U)$ is in the form of  Eq. (\ref{cq}). In this case, $f(V)$ is still  primitive. Thus, if $f(V)=V_A^{'}\otimes V_B^{'}$, then $f(U)$ and $f(V)$ can be discriminated as in subcase (ii-a). If $f(V)=(V_A^{'}\otimes V_B^{'})P$, then it is  easy to discriminate $f(U)$ and $f(V)$, since by letting $|\Phi_X\rangle=f(X)|\alpha\rangle_A|\phi\rangle_B$, we find that $\langle \Phi_U| \Phi_V\rangle=\langle\alpha|V'_A|\phi\rangle\langle\phi|{U'_B}^\dagger V'_B|\alpha\rangle$ which can be zero by choosing  $|\phi\rangle$.

{\noindent\bf Case (iii): Neither $U$ nor $V$ is  primitive, i.e, they are both imprimitive.} Firstly, by
Lemma \ref{Lm2},  we can construct a sequential circuit $f(X)$ consisting of   local unitary  operations  and $X\in\{U,V\}$, such that $f(U)=e^{iu_1\otimes u_2}$ with
$u_1=\sigma_x\oplus 0_{(d_A-2)}$ and $u_2=\sigma_x\oplus 0_{(d_B-2)}$. Thus, $f(U)$
is imprimitive. Now, if $f(V)$ is primitive, then according to case (ii), we know that $f(U)$ and $f(V)$ can be locally
discriminated by a sequential scheme. Otherwise, based on Lemma \ref{Lm3}, we have the following discussion.

Case (iii-a):  $f(V)\neq e^{ixu_1\otimes u_2}$. Let
$F(X)=Wf(X)W^\dagger f(X)$ for  $W\in{\cal S}$. Then in terms of Lemma
\ref{Lm3}, we have $F(U)=I$ and $F(V)\neq I$ for some $W$. Therefore, by
the previous cases, $F(U)$ and $F(V)$ can be locally
discriminated by a sequential scheme, and so for $U$ and $V$.

Case (iii-b): $f(V)= e^{ixu_1\otimes u_2}$. When $x=1$, $f(U)$ and
$f(V)$ are the same and imprimitive. Thus by Lemma \ref{Lm2}, we can
construct a quantum circuit $h(.)$ such that $h(f(U))=U^\dagger$,
and then we have $Uh(f(U))=I$ and $Vh(f(V))=VU^\dagger$.
Therefore, they can be locally discriminated  from the previous
cases. When $x\neq 1$,  discriminating $f(U)$ and $f(V)$ can
be reduced to discriminating $e^{iu_1}$ and $e^{ixu_1}$ as
follows. By inputting $|\varphi\rangle_A|\alpha\rangle_B$ where
$|\alpha\rangle_B$ is an eigenvector of $u_2$ associated with
the eigenvalue $1$, it is easy to check that $e^{ixu_1\otimes
u_2}|\varphi\rangle_A|\alpha\rangle_B=(e^{ixu_1}\otimes
I)|\varphi\rangle_A|\alpha\rangle_B$. Furthermore, we have
\begin{align*}
|\Phi_U\rangle&\equiv f(U)(w_N\otimes I)f(U)\dots(w_1\otimes
I)f(U)|\varphi\rangle_A|\alpha\rangle_B\\&=(e^{iu_1}w_Ne^{iu_1}\dots
w_1e^{iu_1})|\varphi\rangle_A\otimes|\alpha\rangle_B,\\
 |\Phi_V\rangle&\equiv f(V)(w_N\otimes I)f(V)\dots(w_1\otimes
I)f(V)|\varphi\rangle_A|\alpha\rangle_B\\&=(e^{ixu_1}w_Ne^{ixu_1}\dots
w_1e^{ixu_1})|\varphi\rangle_A\otimes|\alpha\rangle_B.
\end{align*}
Therefore, in terms of Lemma \ref{Lm1}, by choosing a suitable input state
$|\varphi\rangle_A$ and auxiliary operations $w_i$, we can get
$|\Phi_U\rangle\perp|\Phi_V\rangle$. Thus, $f(U)$ and $f(V)$ can be
 locally discriminated, and so for $U$ and $V$.

Therefore, we have completed the proof of Theorem \ref{th1}. \qed

\section{Conclusion}\label{Con} A sequential scheme usually represents the most economic strategy for (locally) discriminating two unitary operations.
In this paper we have proved  that any two  bipartite unitary operations $U$ and $V$ acting on $d_A\otimes d_B$ can be locally discriminated by a sequential scheme  without using the inverses of the unitary operations. Compared with the existing related work, the improvement of this paper is twofold. First, the result here applies to the general case of  $d_A\otimes d_B$, whereas Ref. \cite{Li08} only considered the special case of  $d\otimes d$. Second, the sequential scheme here does not use the inverses of  $U$ and $V$, while the inverses were required to construct a sequential scheme in \cite{Li08}. Note that when  $U$ and $V$ are not identified, how to obtain their inverses $U^\dagger$ and $V^\dagger$ is not easy and even not realizable in experiment. Therefore, this paper enhances  the applicability and feasibility of the sequential scheme for locally discriminating unitary operations.

\section*{ACKNOWLEDGMENTS} The author is thankful to the anonymous referees for their valuable comments
and suggestions. This paper is supported in part by the National Natural Science Foundation of China (Nos. 61472452, 61572532) and the National Natural Science Foundation of Guangdong Province of China (No. 2014A030313157), the Science and Technology Program of Guangzhou City of China (No. 201707010194), the Fundamental Research Funds for the Central Universities (Nos. 17lgzd29, 17lgjc24).

\end{document}